\documentclass[aps,pre,twocolumn,showpacs,preprintnumbers,amsmath,amssymb,superscriptaddress]{revtex4}
\usepackage{dcolumn}                    
\usepackage{bm}                        
\usepackage{graphicx}
\usepackage{times}
\usepackage{epstopdf}
\usepackage{color}
\begin{document}

\definecolor{Red}{rgb}{1,0,0}

\definecolor{Blu}{rgb}{0,0,01}

\definecolor{Green}{rgb}{0,1,0}

\newcommand{\red}{\color{Red}}
\newcommand{\blu}{\color{Blu}}
\newcommand{\green}{\color{Green}}

\newcommand{\Imag}{{\Im\mathrm{m}}}   
\newcommand{\Real}{{\mathrm{Re}}}   
\newcommand{\im}{\mathrm{i}}        
\newcommand{\talpha}{\tilde{\alpha}}
\newcommand{\ve}[1]{{\mathbf{#1}}}

\newcommand{\x}{\lambda}  
\newcommand{\y}{\rho}     
\newcommand{\T}{\mathrm{T}}   
\newcommand{\Pv}{\mathcal{P}} 
\newcommand{\vk}{\ve{k}} 
\newcommand{\vp}{\ve{p}} 

\newcommand{\vm}{\boldsymbol{m}} 
\newcommand{\vM}{\boldsymbol{M}}
\newcommand{\X}{\mathcal{X}}
\newcommand{\Hp}{\mathcal{H}_\perp}

\newcommand{\N}{\underline{\mathcal{N}}} 
\newcommand{\Nt}{\underline{\tilde{\mathcal{N}}}} 
\newcommand{\g}{\underline{\gamma}} 
\newcommand{\gt}{\underline{\tilde{\gamma}}} 

\newcommand{\vecr}{\ve{r}} 
\newcommand{\vq}{\ve{q}} 
\newcommand{\ca}[2][]{c_{#2}^{\vphantom{\dagger}#1}} 
\newcommand{\cc}[2][]{c_{#2}^{{\dagger}#1}}          
\newcommand{\da}[2][]{d_{#2}^{\vphantom{\dagger}#1}} 
\newcommand{\dc}[2][]{d_{#2}^{{\dagger}#1}}          
\newcommand{\ga}[2][]{\gamma_{#2}^{\vphantom{\dagger}#1}} 
\newcommand{\gc}[2][]{\gamma_{#2}^{{\dagger}#1}}          
\newcommand{\ea}[2][]{\eta_{#2}^{\vphantom{\dagger}#1}} 
\newcommand{\ec}[2][]{\eta_{#2}^{{\dagger}#1}}          
\newcommand{\su}{\uparrow}    
\newcommand{\sd}{\downarrow}  
\newcommand{\Tkp}[1]{T_{\vk\vp#1}}  
\newcommand{\muone}{\mu^{(1)}}      
\newcommand{\mutwo}{\mu^{(2)}}      
\newcommand{\epsk}{\varepsilon_\vk}
\newcommand{\epsp}{\varepsilon_\vp}
\newcommand{\e}[1]{\mathrm{e}^{#1}}
\newcommand{\dif}{\mathrm{d}} 
\newcommand{\diff}[2]{\frac{\dif #1}{\dif #2}}
\newcommand{\pdiff}[2]{\frac{\partial #1}{\partial #2}}
\newcommand{\mean}[1]{\langle#1\rangle}
\newcommand{\abs}[1]{|#1|}
\newcommand{\abss}[1]{|#1|^2}
\newcommand{\Sk}[1][\vk]{\ve{S}_{#1}}
\newcommand{\pauli}[1][\alpha\beta]{\boldsymbol{\sigma}_{#1}^{\vphantom{\dagger}}}

\newcommand{\eq}{Eq.}
\newcommand{\eqs}{Eqs.}
\newcommand{\cf}{\textit{cf. }}
\newcommand{\ie}{\textit{i.e. }}
\newcommand{\eg}{\textit{e.g. }}
\newcommand{\etal}{\emph{et al.}}
\def\i{\mathrm{i}}

\title{Anomalous Domain Wall Velocity and Walker Breakdown in Hybrid Systems with Anisotropic Exchange}

\author{Henrik Enoksen}
\affiliation{Department of Physics, Norwegian University of
Science and Technology, N-7491 Trondheim, Norway}

\author{Asle Sudb{\o}}
\affiliation{Department of Physics, Norwegian University of
Science and Technology, N-7491 Trondheim, Norway}

\author{Jacob Linder}
\affiliation{Department of Physics, Norwegian University of
Science and Technology, N-7491 Trondheim, Norway}

\date{\today}

\begin{abstract}
It has recently been proposed that spin-transfer torques in magnetic systems with anisotropic exchange can be strongly enhanced, reducing the characteristic current density with up to four orders of magnitude compared to conventional setups. Motivated by this, we analytically solve the equations of motion in a collective-coordinate framework for this type of anisotropic exchange system, to investigate the domain wall dynamics in detail. In particular, we obtain analytical expressions for the maximum attainable domain wall velocity of such a setup and also for the occurrence of Walker breakdown. Surprisingly, we find that, in contrast to the standard case with domain wall motion driven by the non-adiabatic torque, the maximum velocity obtained via the anisotropic exchange torque is completely \textit{independent} of the non-adiabaticity parameter $\beta$, in spite of the torque itself being very large for small $\beta$. Moreover, the Walker breakdown threshold has an opposite dependence on $\beta$ in these two cases, i.e. for the anisotropic exchange torque scenario, the threshold value decreases monotonically 
with $\beta$. These findings are of importance to any practical application of the proposed giant spin-transfer torque in anisotropic exchange systems. 
\end{abstract}
\maketitle

\textit{Introduction}. The concept of spin-transport in magnetic structures  has proven to be of much relevance in terms of both applications and fundamental physics \cite{zutic_rmp_04}. One particularly promising topic in the field of spintronics is electrical control of domain wall motion in textured ferromagnets \cite{grollier_phys_11}. The essential idea is that controllable domain wall motion via spin-transfer torque may be used to represent information. This forms the basis for possible applications such as magnetic random access memory, magnetic racetrack technology, and various types of magnetic logic gates \cite{kent_apl_04, matsunaga_ape_08, parkin_science_08, liu_apl_10}.

In order to make controllable domain wall motion a feasible technology, two key aspects \cite{hayashi_prl_07, pizzini_ape_09} need to be addressed: \textit{i)} the required current density to induce high-speed domain wall motion must be lowered as much as possible and \textit{ii)} the structural deformation of the domain wall triggered at the Walker breakdown \cite{walker} must be delayed as much as possible, allowing for a higher maximum wall velocity. An interesting proposition in the right direction concerning point \textit{i)} was recently made in Ref. \cite{korenev_arxiv_12}. By considering an  {\it anisotropic} exchange interaction of a bilayer system consisting of a ferromagnetic insulator and a semiconducting quantum well, it was proposed that sending a current through the latter part of the system would induce a torque on the texture $\vM(x,t)$ of the ferromagnetic insulator which would be four orders of magnitude stronger than in conventional spin-transfer torque setups. The torque originating with anisotropic exchange was found to be proportional to $1/\beta$ with $\beta$ being the so-called non-adiabaticity parameter \cite{zhang_prl_04}. For the typical non-adiabatic torque term, the corresponding proportionality is $\beta$, which is seen to be much smaller than the torque in the present system with anisotropic exchange, since $\beta \ll 1$. It should be noted that although the term ``non-adiabatic'' for this torque is standard in the literature, it is somewhat misleading since this is a dissipative torque present in the adiabatic limit and not a non-local torque~\cite{garate_prb_09}. However, we will adhere to the established convention in what follows and refer to this term as non-adiabatic.

The usefulness of this novel anisotropic exchange torque in applications relies on the resulting domain wall dynamics, in particular the maximum wall velocity and the occurrence of Walker breakdown. Here, we address these issues by solving analytically the Landau-Lifshitz-Gilbert (LLG) equation in a collective-coordinate framework for the domain wall. The two main degrees of freedom in this treatment are the velocity of the domain wall center $\dot{X}$ and the tilt angle $\phi$ describing the deformation of the domain wall as it propagates. While we confirm the finding of a considerably lowered characteristic current density proposed in Ref. \cite{korenev_arxiv_12}, we find that in contrast to the conventional case with domain-wall motion driven by the non-adiabatic spin-transfer torque, the maximum wall velocity under the influence of an anisotropic exchange torque is completely \textit{independent} of $\beta$. 
  
Deriving analytical expressions both for the maximum velocity and the Walker breakdown threshold, we demonstrate that the latter also behaves differently from the conventional non-adiabatic case. For a system with anisotropic exchange interaction, the breakdown threshold is proportional to $\beta$ and thus decreases as $\beta \to 0$. Finally, we consider the properties of a new hybrid system where the total spin-transfer torque acting on the domain wall has a contribution both from the conventional dissipative torque and the anisotropic exchange torque by means of two separate currents. In this case, we show that the relative magnitude and direction of the currents flowing can be used to tune both the maximum domain wall velocity and the threshold value for Walker breakdown.

\begin{figure}[t!]
\centering
\includegraphics[width=\columnwidth]{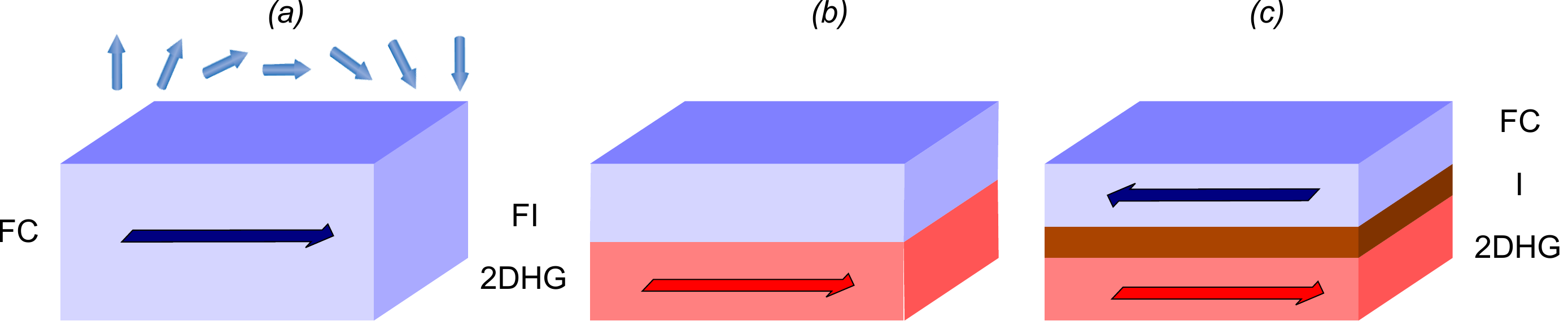}
\caption{(Color online) (a) A conducting ferromagnet (FC) with a N\'eel magnetic domain wall. (b) A bilayer consisting of a ferromagnetic insulator (FI) and a semiconducting quantum well (2DHG). A N\'eel magnetic domain wall makes the ferromagnet textured, and the magnetic coupling between the two layers is strongly anisotropic. (c) The ferromagnet is again conducting and separated from the 2DHG by an insulating layer (I), allowing for separate currents to flow in the FC and 2DHG region. Here, 2DHG denotes a two-dimensional hole gas.}
\label{fig:model}
\end{figure}

\textit{Theory}. The physical systems under consideration are shown in Fig. \ref{fig:model}. In (a), the textured ferromagnet is electrically conducting and corresponds to the conventional case of current-induced domain wall motion. In (b), the ferromagnet is an insulator so that the domain wall motion is induced exclusively via exchange interaction to the semiconducting quantum well when a current passes through the latter. In (c), the ferromagnet is no longer insulating and separate currents may flow in the two layers due to insertion of an electrical insulator between them. Our starting point is to consider a N\'eel-type domain wall $\vM(x,t)$ where the easy (hard) axis of magnetic anisotropy are taken to be along the $\hat{z}$ ($\hat{y}$) direction \cite{kohno_jpsj_06}
\begin{align}
\vm(x,t) = [\sin\theta(x)\cos\phi(t), \sin\theta(x)\sin\phi(t),\sigma \cos\theta(x)],
\end{align}
where $\sigma$ denotes the topological charge of the domain wall. Here, we have defined the magnetization unit vector as $\vm(x,t) = \vM(x,t)/M_0$ with $M_0 = |\vM(x,t)|$ being the saturation magnetization. The tilt angle $\phi(t)$ represents a deformation mode of the domain wall from its equilibrium configuration, while $\theta(x)$ represents the angle between the magnetization and the easy axis. The domain wall texture in the static case is determined by the exchange stiffness and magnetic anisotropy axes present in the system. We emphasize that we have also performed the calculations to be presented for a Bloch-type domain wall profile (same easy axis, but hard axis along the $\hat{x}$ direction) and find identical results. This equivalence between the wall profiles no longer holds in the presence of spin-orbit interactions \cite{ryu, linder_prb_13, fert_arxiv_13}. We do not consider spin-orbit interactions here, as it is not central to the main results. The interactions in our system are taken into account via the effective field $\boldsymbol{H}_\text{eff}$
\begin{align}\label{eq:heff}
\boldsymbol{H}_\text{eff} &= \frac{2 A_\text{ex}}{M_0}\nabla^2\boldsymbol{m} - H_\perp m_y \hat{y} + H_km_z\hat{z} + \boldsymbol{H}_\text{ext}.
\end{align}
Here, $A_{\text{ex}}$ is the exchange coupling constant, $H_k$ and $H_{\perp}$ are the anisotropy fields for the easy and hard axes, respectively, and $\boldsymbol{H}_{\text{ext}}$ is an external magnetic field. The components of the magnetization vector depend on both space and time according to: $\cos\theta = \tanh\Big(\frac{x-X(t)}{\lambda}\Big),$ $\sin\theta = \text{sech}\Big(\frac{x-X(t)}{\lambda}\Big).$ Here, $\lambda = \sqrt{2A_{\text{ex}}/M_0 H_k}$ is the domain wall width and $X(t)$ is the position of the domain wall center. Assuming that the easy axis anisotropy field $H_k$ is larger than its hard axis equivalent $H_\perp$, i.e. $|H_k|\gg|H_\perp|$, the domain wall may be treated as rigid considering $\phi(t)$ as the only relevant deformation mode of the wall. 

To account for the novel anisotropy exchange torque, one must add an extra term which was derived in Ref. \cite{korenev_arxiv_12}, to the LLG-equation. The physical setup is as follows: consider the exchange interaction between the quantum well and the ferromagnetic film in Fig. \ref{fig:model}(b). Assuming a $p$-type hole gas, the exchange arises from the overlap of wavefunctions between the holes in the quantum well and the atoms in the ferromagnet. The point is now that this exchange interaction is strongly anisotropic if only the heavy-hole 
subband is filled and the splitting to the light-hole band exceeds the exchange splitting $J$ \cite{merkulov_prb_95}. In effect, we are considering an exchange interaction of the 
type $-J S_z(\vecr)m_z(\vecr)$ where $S_z$ and $m_z$ are the $z$-components of the spin densities of heavy holes and the magnetization unit vector, respectively. One finds that the 
full equation which governs the magnetization dynamics reads \cite{korenev_arxiv_12}
\begin{align}\label{eq:LLG}
\partial_t \vm &= -\tilde{\gamma} \vm \times \boldsymbol{H}_\text{eff} + \tilde{\alpha} \vm \times \partial_t \vm + \boldsymbol{\tau}_\text{QW}, \notag\\
\boldsymbol{\tau}_\text{QW} &= \frac{v_\text{s,QW}}{\beta_\text{QW}} (\vm\times\hat{z})(\partial_x m_z).
\end{align}
Here, $\tilde{\gamma}$ and $\tilde{\alpha}$ are the renormalized gyromagnetic ratio and Gilbert damping constant, respectively, whereas $\beta_\text{QW}$ is the non-adiabaticity parameter proportional to the spin relaxation time $\tau_\text{QW}^{-1}$ of itinerant holes in the quantum well. The origin and value of this parameter is debated in the literature, although there appears to be consensus that spin-relaxation processes (which may be model-dependent) are essential \cite{kohno_jpsj_06, tserkovnyak_prb_06, garate_prb_09}. 
We have also introduced the spin velocity $v_\text{s,QW}$ of the current in the semiconducting layer, which is proportional to the current density (see below). We will obtain the equations of motion for the collective coordinates $\{X,\phi\}$ and solve these analytically, thus obtaining a description of how the domain wall velocity $v_\text{DW} \equiv \dot{X}$ and the Walker breakdown criterion $\dot{\phi}\neq0$ depend on the external parameters of the system, such as the applied current. The breakdown threshold is the critical current density where the domain wall starts deforming from its original shape rather than simply being translated along the current direction. 

The theory up to now has been described with the setup in Fig. \ref{fig:model}(b) in mind. For comparison, we will also consider the system depicted in Fig. \ref{fig:model}(a) \cite{zhang_prl_04, thiaville_epl_05}. The current flowing through the textured ferromagnet generates two different types of torque terms compared to the anisotropic exchange mechanism, in the literature often referred to as the adiabatic and non-adiabatic torque. Their combined effect is captured in a term $\boldsymbol{\tau}_\text{FM}$ which is added to the right hand side of Eq. (\ref{eq:LLG}):
\begin{align}
\boldsymbol{\tau}_\text{FM} = v_\text{s,FM}\partial_x\vm - \beta_\text{FM} v_\text{s,FM}\vm\times \partial_x\vm.
\end{align}
This torque term is controlled by the spin current $v_\text{s,FM}$ flowing in the ferromagnetic conductor, and the non-adiabiticity parameter $\beta_\text{FM}$ is in general 
different from the one in the quantum well, $\beta_\text{QW}$. Finally, in Fig. \ref{fig:model}(c) we consider a scenario where both the ferromagnet and quantum well are 
conducting with an insulator dividing the two regions and thus permitting two separate currents to flow in each layer. In this case, both $\boldsymbol{\tau}_\text{QW}$ 
and $\boldsymbol{\tau}_\text{FM}$ should be considered simultaneously in the LLG equation.

\textit{Results and Discussion}. For a system where all torque terms are present, like the one depicted in Fig.~\ref{fig:model}(c), the equations of motion for $\{X,\phi\}$ are
\begin{align}
  \sigma \left(1 + \tilde{\alpha}^2\right)\dot{\tilde{X}} &= \sin{2\phi} -\sigma\left(1+\tilde{\alpha}\beta_{\text{FM}}\right) \tilde{v}_{\text{s,FM}} +\sigma\frac{\tilde{\alpha} \tilde{v}_{\text{s,QW}}}{2\beta_{\text{QW}}}, \label{eq:eqmox}  \\
  \left(1 + \tilde{\alpha}^2\right)\dot{\tilde{\phi}} &= -\tilde{\alpha}\sin{2\phi} + \sigma\left(\tilde{\alpha} - \beta_{\text{FM}}\right) \tilde{v}_{\text{s,FM}} + \sigma\frac{\tilde{v}_{\text{s,QW}}}{2\beta_{\text{QW}}}. \label{eq:eqmophi}
\end{align}
We will consider the general form for the equations of motion to begin with, whence we obtain the cases shown in Fig. \ref{fig:model}(a) and (b) by setting some parameters equal to zero in the final result for the domain wall velocity and Walker breakdown threshold. Here, we have introduced the dimensionless parameters $\dot{\tilde{X}} = \partial\left(X/\lambda \right)/\partial\tilde{t}$, $\dot{\tilde{\phi}} = \partial\phi/\partial\tilde{t}$, $\tilde{t} = (\tilde{\gamma}H_{\perp}/2)t$, and $\tilde{v} = \left(2/\tilde{\gamma}\lambda H_{\perp}\right)v$.  We have used $v_{\text{s,FM}} = \left(\hbar\tilde{\gamma}/2eM_0\right)Pj$ where $e$ is the electronic charge, $P$ is the spin polarization of the current, and $j$ is the current density. Parameter values are $P = 0.7$, $M_0 = 5\cdot10^5 \text{A/m}$, $A_\text{ex} = 10^{-11} \text{J/m}$, $H_{\perp} = 0.04 \text{T}$ and $H_k = 0.4 \text{T}$. By solving Eq.~\eqref{eq:eqmophi} analytically and inserting the solution into \eq\eqref{eq:eqmox}, we obtain an expression for the average domain wall velocity 
\begin{equation}
  \langle \dot{\tilde{X}} \rangle = -\frac{\beta_{\text{FM}}}{\talpha}\tilde{v}_{\text{s,FM}} + \frac{1}{2\talpha\beta_{\text{QW}}}\tilde{v}_{\text{s,QW}} - \frac{\mathrm{sgn}\left(\tilde{J}\right)}{1+\talpha^2}\sqrt{\tilde{J}^2-1},
  \label{eq:avgdwvelocity}
\end{equation}
where $\tilde{J} = \left(1-\frac{\beta_{\text{FM}}}{\talpha}\right)\tilde{v}_{\text{s,FM}} + \frac{1}{2\talpha\beta_{\text{QW}}}\tilde{v}_{\text{s,QW}}$. Details on this procedure can be found in \eg\cite{tatara_physrep_08}. 

As stated above, Walker breakdown occurs when $\dot{\phi} \neq 0$. From \eq\eqref{eq:eqmophi} we obtain, for a constant tilt angle, the result
\begin{equation}
  \sin(2\phi) = \sigma\left(1-\frac{\beta_{\text{FM}}}{\talpha}\right)\tilde{v}_{\text{s,FM}} + \sigma\frac{\tilde{v}_{\text{s,QW}}}{2\talpha\beta_{\text{QW}}},
  \label{eq:walkerbreakdowncondition}
\end{equation}
which gives the domain wall velocity
\begin{equation}\label{eq:h2}
  \langle \dot{\tilde{X}} \rangle = -\frac{\beta_{\text{FM}}}{\talpha}\tilde{v}_{\text{s,FM}} + \frac{1}{2\talpha\beta_{\text{QW}}}\tilde{v}_{\text{s,QW}}.
\end{equation}
This is the same as the first two terms in \eq\eqref{eq:avgdwvelocity}. We see that the domain wall velocity reaches a maximum when the two currents flow in opposite directions. 
However, when the right hand side of \eq\eqref{eq:walkerbreakdowncondition} becomes greater than unity, domain wall deformation sets in. 

Consider first the results for the case shown in Fig. \ref{fig:model}(a) \cite{zhang_prl_04, thiaville_epl_05}. Setting $\tilde{v}_\text{s,QW}=0$ in the above results, one finds that Walker breakdown sets in when $\tilde{v}_{\text{s,FM}c} = \frac{1}{1-{\beta_\text{FM}/\talpha}}$ and the maximum domain wall velocity attainable in this setup is $\langle \dot{\tilde{X}}\rangle_c = \frac{1}{|1 - \talpha/\beta_\text{FM}|}$. We emphasize that both $\tilde{v}_{\text{s,FM}c}$ and $\langle \dot{\tilde{X}}\rangle_c$ are normalized and thus dimensionless quantities, to facilitate comparison between the different setups in Fig. \ref{fig:model}. Importantly, the normalization constants are independent of $\talpha$ and $\beta$. 
The above results  show that when $\talpha = \beta_\text{FM}$, Walker breakdown is absent and the maximum domain wall velocity has no upper bound - it is simply proportional to the applied current density. In practice, however, these parameters cannot be controlled or tuned in a well-defined manner such that this limit is not easily obtained. A notable feature is that for large $\beta_\text{FM}$, the threshold current is lowered whereas the maximum wall velocity increases. 

With this in mind, we analyze the results for domain wall dynamics in the anisotropic exchange system Fig. \ref{fig:model}(b). By using Eqs. (\ref{eq:avgdwvelocity})-(\ref{eq:h2}), we find after some calculations the following results for Walker breakdown and maximum domain wall velocity
\begin{align}\label{eq:10}
\tilde{v}_{\text{s,QW}c} &= 2\talpha\beta_\text{QW},\notag\\
\langle \dot{\tilde{X}}\rangle_c &= 1.
\end{align}
We note that, despite the fact that the spin-transfer torque is increased by as much as four orders of magnitude in the anisostropic exchange system, due to the 
coefficient $1/\beta_\text{QW}$, the maximum domain wall velocity is \textit{completely independent} of the value of $\beta_\text{QW}$. This should be compared 
to the standard setup in Fig. \ref{fig:model}(a), where the wall is driven by the non-adiabatic spin-transfer torque, yielding a maximum domain wall velocity 
strongly dependent on the value of $\beta_{\text{FM}}$. The above result may be understood as a result of competition between the magnitude of the spin-transfer 
torque and the occurrence of Walker breakdown. Looking at Eq. (\ref{eq:h2}), it is seen that the domain wall velocity is related to the spin velocity 
$\tilde{v}_{\text{s,QW}}$ via a constant of proportionality $\sim\beta_{\text{QW}}^{-1}$. At the same time, the critical value of the spin velocity where 
the domain wall no longer is stable towards deformation is given by $\tilde{v}_{\text{s,QW}c}$ in Eq. (\ref{eq:10}) and is proportional to $\beta_{\text{QW}}$. 
As a result, the maximum domain wall velocity (which occurs right at the critical value for $\tilde{v}_{\text{s,QW}}$) is independent of $\beta_\text{QW}$, 
since the dependence on this parameter cancels out when multiplying the domain wall velocity with the critical spin velocity. Physically, this means 
that although the very large magnitude of the torque gives rise to a rapid increase of wall velocity with current, the same property of the torque also 
renders the wall unstable towards deformation faster than in the conventional case. The dependence on $\beta$ for these two effects is such that they 
exactly compensate.

\begin{figure}[t!]
  \centering
  \includegraphics[width=\columnwidth]{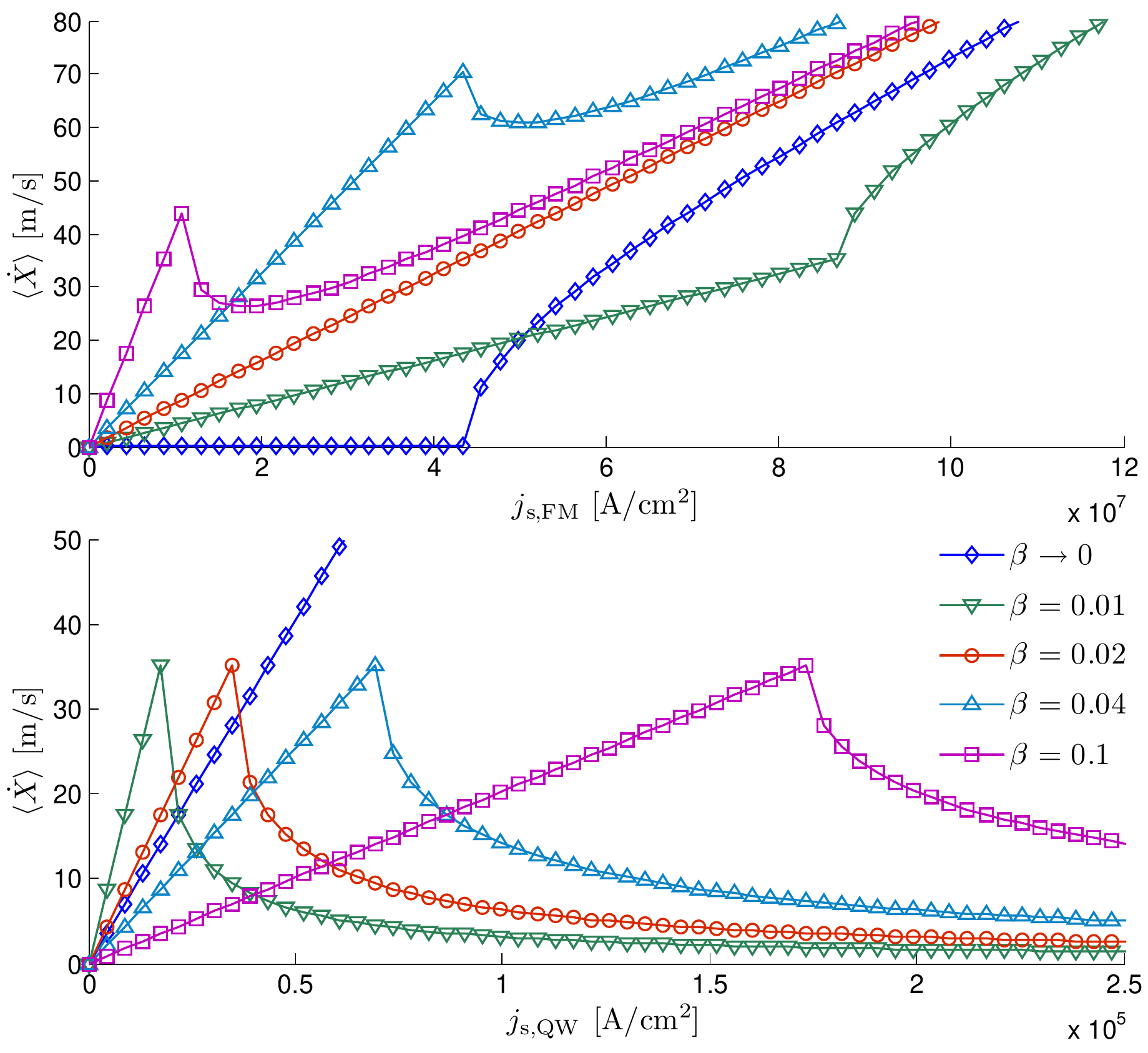}
  \caption{(Color online) Average domain wall velocity obtained for the setups in Fig. \ref{fig:model}(a) (corresponding to the top panel) and (b) (corresponding to the bottom panel),
 for several choices of $\beta$. We have fixed $\talpha=0.02$. Note the difference in order of magnitude on the $x$-axes, demonstrating the vastly different current 
magnitudes required. For $\beta\to0$ in the bottom panel, Walker breakdown occurs almost immediately (not seen) and the increasing domain wall velocity is above 
threshold so that the wall is continuously deforming. The maximum wall velocity prior to Walker breakdown is seen to be independent of $\beta$ in the bottom 
panel [Fig. \ref{fig:model}(b)] in contrast to the top panel [Fig. \ref{fig:model}(a)].  }
  \label{fig:avgdwvelocity}
\end{figure}

Another qualitatively new aspect of the anisotropic exchange case is in the manifestation of Walker breakdown. The situation is different from the standard setup in 
Fig. \ref{fig:model}(a). Namely, the critical current where Walker breakdown takes place now \textit{increases} with $\beta_\text{QW}$. Effectively, this means that the Walker 
breakdown threshold decreases as the spin-transfer torque increases, yet the maximum wall velocity is unaffected by any change in $\beta_\text{QW}$. This is a unique 
feature of the spin-transfer torque originating with anisotropic exchange. The various domain wall dynamics is depicted in Fig. \ref{fig:avgdwvelocity}, 
comparing the setups in Figs.~\ref{fig:model}(a) and.~\ref{fig:model}(b).

The domain-wall velocity resulting from $\boldsymbol{\tau}_\text{QW}$ reaches appreciable values at much lower current densities compared to a domain wall driven by $\boldsymbol{\tau}_\text{FM}$. At the same time, the maximum attainable wall velocity is independent of $\beta_{\text{QW}}$, while the maximum velocity obtained via $\boldsymbol{\tau}_\text{FM}$ can be very high for large values of $\beta_{\text{FM}}$. An intriguing scenario could be realized if one were to combine these two torques to act 
simultaneously on the same domain wall. This can be realized experimentally as shown in Fig. \ref{fig:model}(c), where two separate currents flowing in the FM and QW layers control the magnitude of these torques. Introducing the ratio between the spin currents in the ferromagnet and the semiconducting quantum well, \ie $a = v_{\text{s,QW}}/v_{\text{s,FM}} $, we can express $\tilde{J}$ from \eq\eqref{eq:avgdwvelocity} as
\begin{equation}
  \tilde{J} = \tilde{v}_{\mathrm{s,FM}}\left(\left(1-\frac{\beta_\text{FM}}{\talpha}\right)b + \frac{a}{2\talpha\beta_\text{QW}} \right),
\end{equation}
where $b$ is a parameter which controls the direction of $\tilde{v}_{\text{s,FM}}$: $b=$ 1 (-1) for parallell (antiparallell) flow of the currents. Note that the currents in the QW and FM region can be set to zero respectively by $a=0$ or $b=0$.

Since the Walker breakdown condition is $\tilde{J}^2 > 1$, we can now identify the critical spin current velocity and the corresponding critical domain wall velocity for the setup in Fig. \ref{fig:model}(c) where both $\boldsymbol{\tau}_\text{QW}$ and $\boldsymbol{\tau}_\text{FM}$ are active
\begin{align}
  \tilde{v}_{\text{s,FM}c} &= \frac{1}{\left(1-\frac{\beta_\text{FM}}{\talpha}\right)b + \frac{a}{2\talpha\beta_\text{QW}}}, \label{eq:critcurrent}\\
  \langle \dot{\tilde{X}}\rangle_c &= \frac{1}{1 - \frac{\talpha b}{\beta_\text{FM} b - \frac{a}{2\beta_\text{QW}}}}. \label{eq:critvelocity}
\end{align}
In this way, one can control both the maximum attainable domain wall velocity and the occurrence of Walker breakdown by the relative magnitude $a$ and relative sign $b$ of the currents flowing in the FM and QW layer. It should be kept in mind that the analysis performed here is for an idealized domain wall system without any defects or pinning potentials. Nevertheless, our results demonstrate the qualitatively different role played by the non-adiabaticity parameter $\beta$ in the anisotropic exchange system.

\textit{Summary}. In summary, we have shown that although the spin-transfer torque in magnetic systems with anisotropic exchange can be made much larger than the conventional non-adiabatic torque, due to the former being proportional to $\beta^{-1}$ rather than $\beta$, the maximum domain wall velocity is \textit{independent} of $\beta$ in contrast to what one might expect. The Walker breakdown threshold decreases monotonically with $\beta$, which also differs from the conventional scenario. These findings are of practical relevance to any application of the proposed giant spin-transfer torque in anisotropic exchange systems. 

\textit{Acknowledgments.} H. E. acknowledges support from NTNU. J.L. and A.S. acknowledge support from the Research Council of Norway through Grant Nos 205591/V20 and
 216700/F20. A.S. acknowledges useful discussions with F. S. Nogueira.

\end{document}